\documentclass{ws-p10x7}

\begin{document}

\title{Heavy Quark Production in $\gamma\gamma$ Collisions}

\author{A.~B\"ohrer}

\address{Fachbereich Physik, Universit\"at Siegen, D-57068 Siegen, 
Germany\\E-mail: armin.boehrer@cern.ch}

\twocolumn[\maketitle\abstract{
New results on inclusive heavy quark production in $\gamma\gamma$ 
collisions are presented. Charm and bottom production are investigated 
at LEP~II energies 
by the experiments ALEPH, DELPHI, L3, and OPAL. The total and differential 
cross sections for charm quarks are measured. The contributions from the 
direct and single-resolved processes are separated and their fractions 
quantified. More detailed studies, such as the dependence of the 
cross section on the two-photon centre-of-mass energy and the charm structure 
function $F^2_{\gamma,{\mathrm {c}}}$, are reported. The inclusive bottom 
cross section is presented. Measurements are compared to next-to-leading 
order calculations.}]

\section{Introduction}

\begin{picture}(10,0)(0,0)
\put(410.,220.){SI-2000-8}
\put(410.,210.){August 2000}
\put(0.,-455.){Invited talk given at the ICHEP'2000, Osaka.}
\end{picture}
The production of heavy flavour in two-photon collisions is dominated 
by two processes, the direct and the single-resolved process. Both contribute 
in equal shares to heavy flavour final states at 
LEP~II energies. The large quark 
mass allows reliable perturbative calculations for the direct contribution. 
The single-resolved one in addition depends on the gluon density of the 
photon.

In this article the new measurements\cite{all} from the LEP experiments 
on heavy flavour production and the comparison with next-to-leading order 
calculations\cite{nlo} are summarized. Only the main results are given. For 
details the reader is referred to the original papers.

\section{Heavy Quark Identification}

A clear, non-ambiguous signal for charm quark production is the presence of 
a ${\mathrm{D}}^{*+}$. This gold-plated signature has been exploited 
by all four LEP collaborations. 
Both the production probability ${\cal{P}}
({\mathrm{c}} \rightarrow {\mathrm{D}}^{*+})$ and the branching ratio to 
${\mathrm{D}}^{0}\pi^+$ as well as the ${\mathrm{D}}^{0}$ branching 
ratios are reasonably well-known. 

The leptonic decay has been used by ALEPH ($\mu^{\pm}$ for charm) and L3 
($\mu^{\pm}$ and ${\mathrm{e}}^{\pm}$ both for charm and bottom). Though 
these analyses rely on the momentum spectrum as theoretical input, the 
statistics is substantially increased.

\begin{figure}
\epsfxsize\columnwidth
\figurebox{\columnwidth}{315pt}{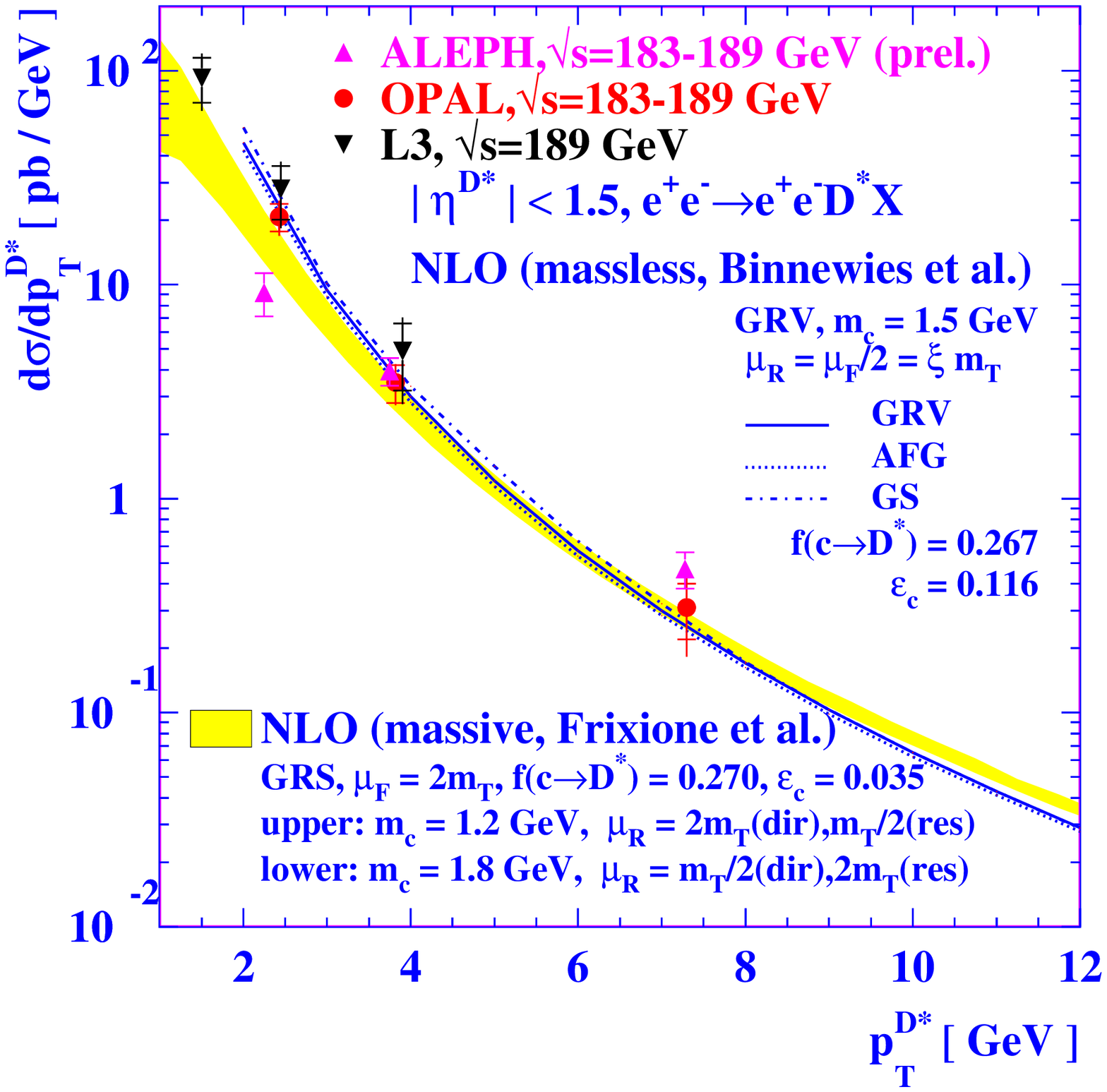}
\caption{Differential cross section ${\mathrm{d}}\sigma / 
{\mathrm{d}}p_{\mathrm T}^{\mathrm {D*}}$ versus the transverse momentum 
$p_{\mathrm T}^{\mathrm {D*}}$ of the ${\mathrm{D}}^{*+}$.}
\label{fig:diffpt}
\end{figure}

\section{Differential Charm Cross Section}

The good detector resolution and with the low kinetic energy available, the 
${\mathrm{D}}^{*+}$ method allows for a measurement of the differential 
cross section in transverse momentum (see Figure~\ref{fig:diffpt}) and 
pseudorapidity. The three experiments, ALEPH, L3, OPAL, 
agree among themselves and with the theoretical expectation, that 
the distribution in pseudorapidity is about constant. 
In transverse momentum the massless approach follows the OPAL measurements, 
while the massive calculation is closer to the ALEPH data.

\section{Charmed Particle Production}

Charmed particles  other than ${\mathrm{D}}^{*+}$ have only been measured 
by DELPHI. From the extracted signal and the efficiencies, the determined 
charm cross section for ${\mathrm{D}}^{\pm}$, 
${\mathrm{D}}^{0}$, and $\Lambda_{\mathrm{c}}$ agree 
with the expectation, which include both direct and resolved processes. 
They are also within the errors consistent with the (2J+1)-relation 
$\sigma_{\mathrm{D}^{*+}} = 3 \cdot \sigma_{\mathrm{D}^0} = 
3 \cdot \sigma_{\mathrm{D}^+}$.

\begin{figure}
\epsfxsize\columnwidth
\figurebox{\columnwidth}{315pt}{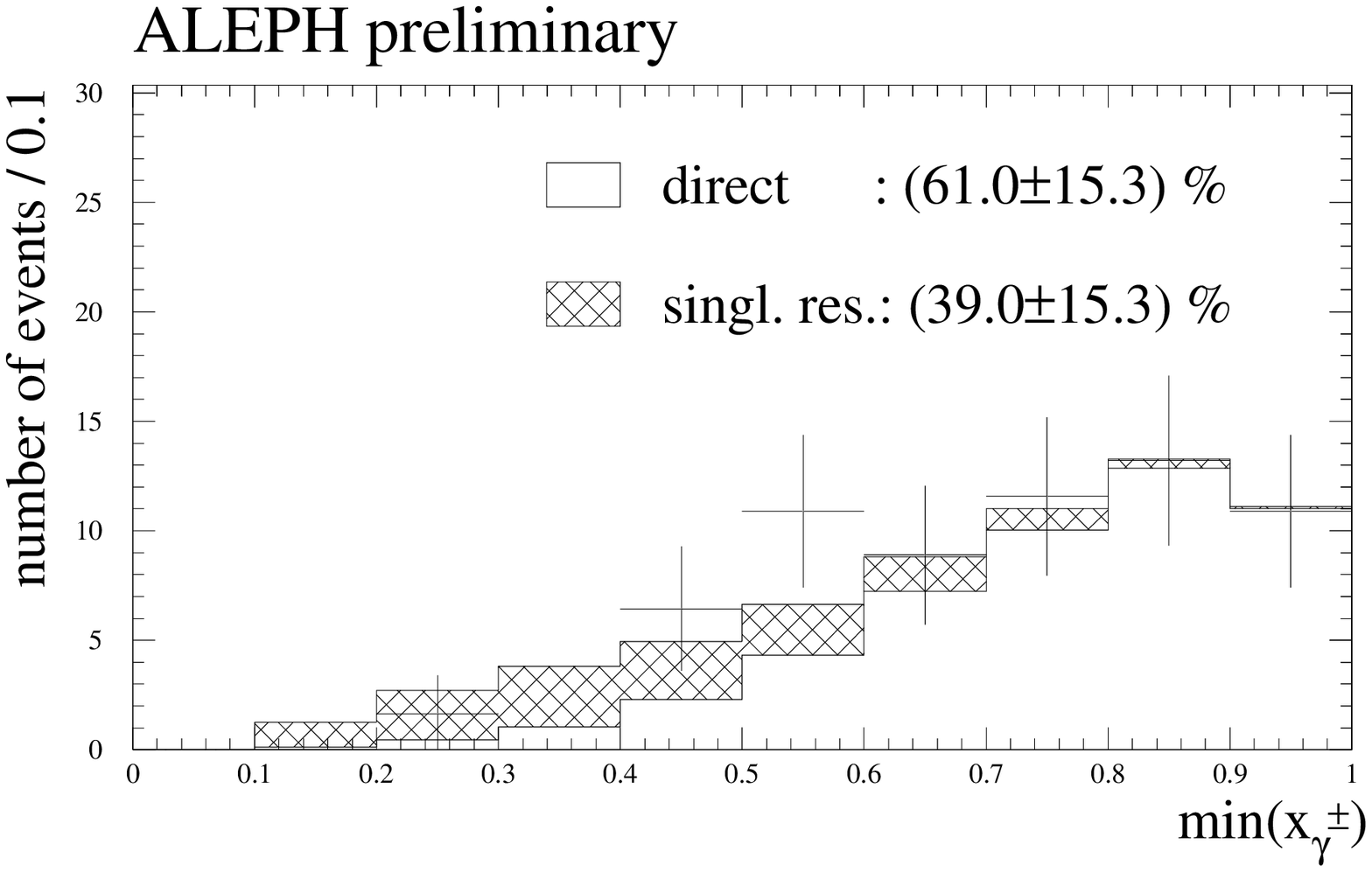}
\caption{Differential cross section ${\mathrm{d}}\sigma / 
{\mathrm{d}}p_{\mathrm T}^{\mathrm {D*}}$ versus the transverse momentum 
$p_{\mathrm T}^{\mathrm {D*}}$ of the ${\mathrm{D}}^{*+}$.}
\label{fig:fdirsr}
\end{figure}

\section{Fraction of Direct and Single-Resolved Contributions}

Direct and single-resolved contribution can be separated using the fact that 
in the resolved one the remnant jet carries away a part of the invariant 
mass available in the $\gamma\gamma$ collision. Two variables have been 
used: 1) $x_{\mathrm T} = p_{\mathrm T}^{\mathrm{D*}} / W_{\mathrm{vis}}$, 
which is the ratio of $p_{\mathrm T}^{\mathrm{D*}}$, a measure for the 
invariant mass of the 
${\mathrm{c}}\bar{{\mathrm{c}}}$ system, and the visible invariant mass 
$W_{\mathrm{vis}}$, a measure for the invariant mass of the $\gamma\gamma$ 
system; 2) $x^{\mathrm{min}}_{\gamma}$ (see Figure~\ref{fig:fdirsr}), the 
minimum of $x^{\pm}_{\gamma} = \sum_{\mathrm{jets}} (E \pm p_z) / 
\sum_{\mathrm{part}} (E \pm p_z)$, a measure 
for the fraction of particles, which do not escape in the remnant jet.

ALEPH measures a relative contribution $r_{\mathrm{dir}} : r_{\mathrm{res}} 
= 62 : 38$ in their acceptance range in agreement with the NLO by Frixione 
($70 : 30$). OPAL obtains $51 :49$.

\section{Charm Cross Section as function of $W_{\gamma\gamma}$}

The L3 collaboration has measured the charm cross section as 
function of the two-photon centre-of-mass energy. The charm-flavoured 
quarks are identified by their semi-leptonic decays to electrons. 
A parameterization of the 
form $\sigma_{\mathrm{tot}} = A s^{\varepsilon} + Bs^{-\eta}$ (Pomeron $+$ 
Reggeon) describes the data well. The PYTHIA Monte Carlo clearly fails, 
predicting only $66\%$ of the total cross section. This may be 
partially attributed to next-to-leading order corrections, which are not 
included in PYTHIA. The Pomeron slope fitted from the data is steeper than 
the rise observed in $\sigma(\gamma\gamma \rightarrow 
{\mathrm{q}}\bar{{\mathrm{q}}}X)$.

\section{Charm Structure Function $F^2_{\gamma,{\mathrm {c}}}$}

\begin{figure}
\epsfxsize\columnwidth
\figurebox{\columnwidth}{150pt}{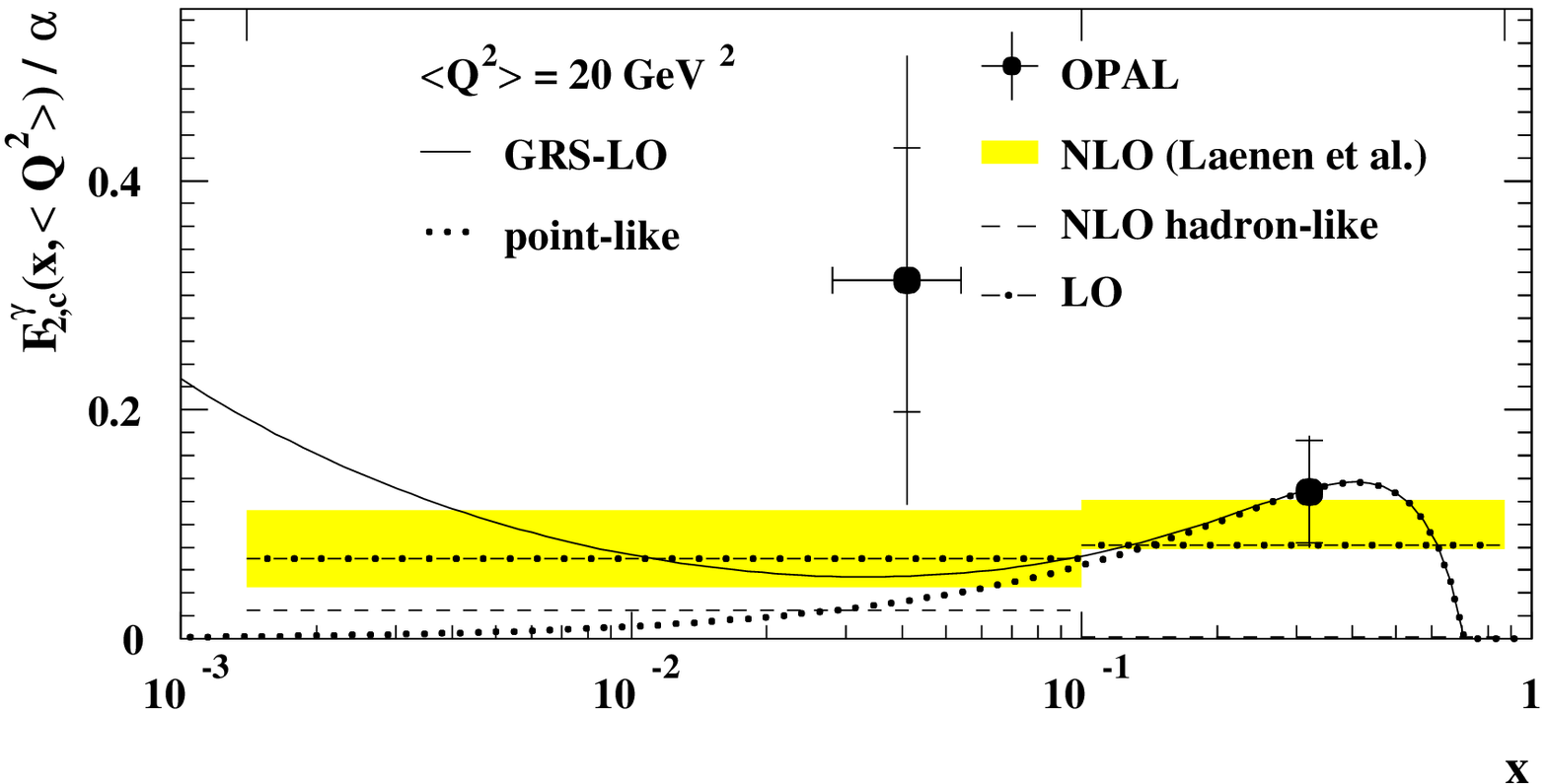}
\caption{Charm structure function obtained with single-tagged events.}
\label{fig:f2gamc}
\end{figure}

When one of the scattered beam particles is detected, the event can be used 
to determine the charm structure function $F^2_{\gamma,{\mathrm {c}}}$. With 
30 such single-tagged events with a ${\mathrm{D}}^{*+}$ meson, the OPAL 
collaboration performed a first measurement. (See the article of 
S.~Soeldner-Rembold in this proceedings for a general overview on 
photon structure functions.) The HERWIG and the Vermaseren Monte Carlo, 
which nicely describe the data, are used for unfolding. The result is 
displayed in Figure~\ref{fig:f2gamc}. The comparison with the calculations 
shows that a point-like contribution is not sufficient to describe the data. 
A hadron-like part is needed. The data even exceed the models, though the 
measurement errors are still too large to be conclusive.

\section{Bottom Quark Production}

Bottom production is measured by the L3 collaboration using the 
fact, that the momentum as well as the transverse momentum of leptons 
with respect to the closest jet is higher for muons and electrons from bottom 
than for background, which is mainly charm. Events with bottom quarks are 
selected with an efficiency of $1.2\%$ and $1.0\%$ for muons and 
electrons, resp. The purity is about $50\%$. The cross section is measured 
to be about three times the prediction from NLO calculations.

\section{Inclusive Charm and Bottom Cross Section}

\begin{figure}
\epsfxsize\columnwidth
\figurebox{\columnwidth}{315pt}{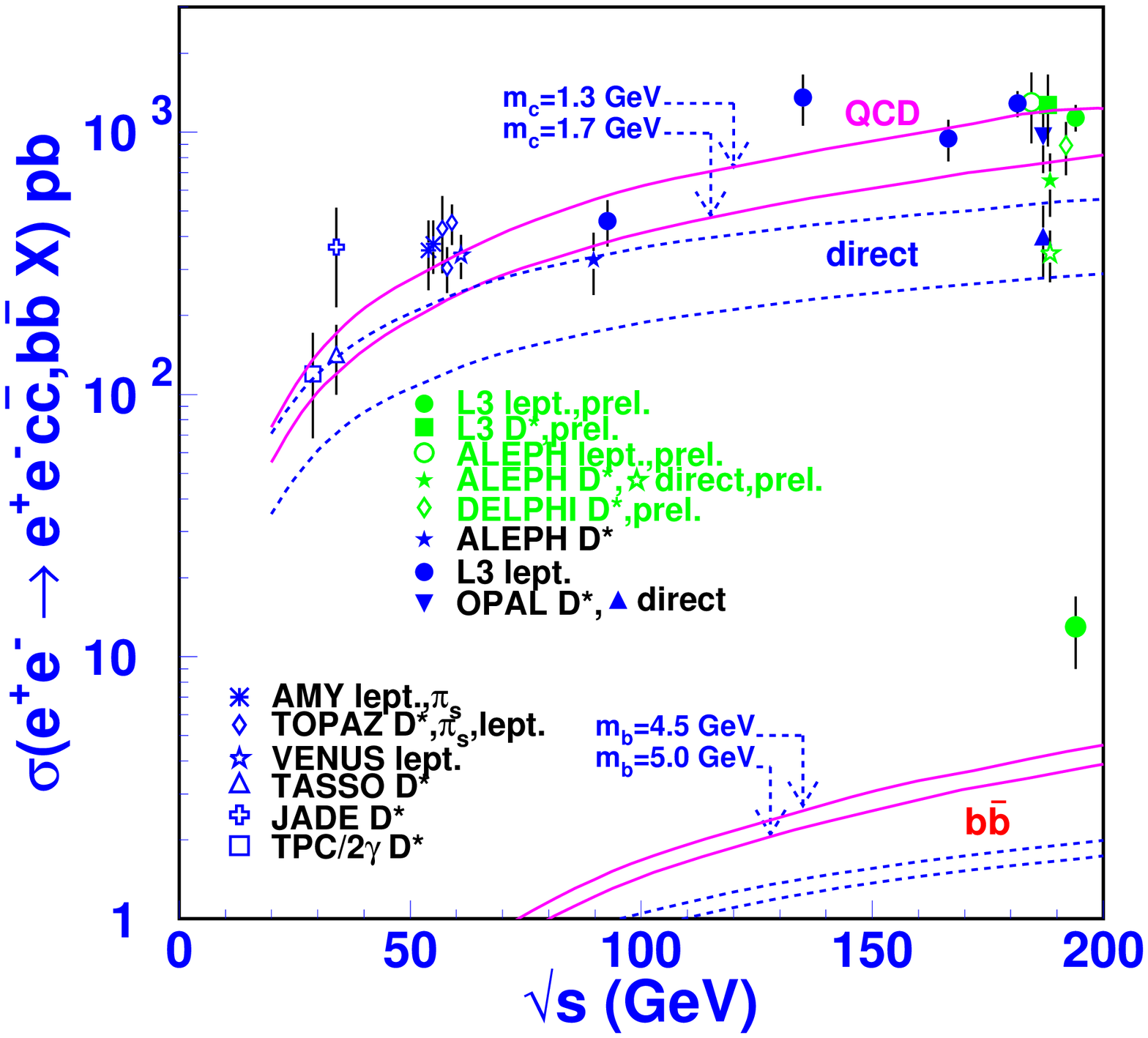}
\caption{Charm and bottom quark production in $\gamma\gamma$ collisions 
as a function of the ${\mathrm{e}^+}{\mathrm{e}^-}$ centre-of-mass energy. 
The predictions calculated in NLO are also given as solid lines. The dashed 
lines represent the contribution from the direct process only.}
\label{fig:iccbbx}
\end{figure}

The total cross section measurements are summarized in Figure~\ref{fig:iccbbx}. 
The results are compared to NLO calculations. Previously published results 
from LEP and those from lower energy colliders are also given.

\section{Summary}

The four LEP experiments have provided good measurements of the 
heavy quark production in $\gamma\gamma$ collisions. The inclusive 
measurement of charm is in agreement with QCD prediction with clear 
evidence for the gluon content in the photon. Next-to-leading order 
contributions seem important. The agreement among the measurements 
of the experiments is fair; comparing the various analysis techniques, 
D-meson versus lepton, the latter tends to give higher cross sections. 
The charm studies are at a transition to precision measurements: 
detailed investigations such as direct/single-resolved, 
$\sigma(W_{\gamma\gamma})$, $F^2_{\gamma,{\mathrm {c}}}$ are performed. 
The bottom production cross section is predicted too low.
 
\section*{Acknowledgements}

I thank S.~Soeldner-Rembold and V.~Andreev, who provided 
Figures~\ref{fig:diffpt} and \ref{fig:iccbbx}.

\end{document}